\documentclass[aps,prb,twocolumn,showpacs,amsmath,amssymb,letterpaper, floatfix, 10pt]{revtex4-1}
\usepackage{graphicx} %
\usepackage{dcolumn} 
\usepackage{bm} 
\usepackage{relsize}
\usepackage{amsmath}
\usepackage{amssymb}
\usepackage{color}
\usepackage[
pdfstartview=XYZ,
bookmarks=false,
colorlinks=true,
linkcolor=blue,
urlcolor=blue,
citecolor=blue,
bookmarks=false,
linktocpage=true, 
hyperindex=false
]{hyperref}
\usepackage{psfrag}
\usepackage[integrals]{wasysym}
\usepackage{epstopdf}
\usepackage{tocloft}
\usepackage{tikz}
\usepackage[normalem]{ulem}
\usepackage{hyphenat}

\usetikzlibrary{shapes}
\usetikzlibrary{arrows, decorations.markings}

\tikzstyle{vecArrow} = [thick, red, decoration={markings,mark=at position
   1 with {\arrow[semithick, red]{open triangle 60}}},
   double distance=1.4pt, shorten >= 4.5pt,
   preaction = {decorate},
   postaction = {draw,line width=1.4pt, white,shorten >= 4.5pt}]
\tikzstyle{innerWhite} = [semithick, white,line width=1.4pt, shorten >= 4.5pt]

\tikzstyle{decision} = [diamond, draw, fill=blue!10, 
    text width=4.5em, text badly centered, node distance=2.75cm, inner sep=0pt, minimum width=4.5em]
\tikzstyle{block} = [rectangle, draw, fill=blue!10, 
    text width=5em, text centered, rounded corners, minimum height=3em]
\tikzstyle{line} = [draw, -latex']

\begin{document}

\title{Self-consistent model of spin accumulation magnetoresistance in ferromagnet-insulator-semiconductor tunnel junctions}
\author{Ian Appelbaum}
\email{appelbaum@physics.umd.edu}
\author{Holly N. Tinkey}
\author{Pengke Li}
\affiliation{Department of Physics and Center for Nanophysics and Advanced Materials, U. Maryland, College Park, MD 20742}

\begin{abstract}
Spin accumulation in a paramagnetic semiconductor due to voltage-biased current tunneling from a polarized ferromagnet is experimentally manifest as a small additional spin-dependent resistance. We describe a rigorous model incorporating the necessary self-consistency between electrochemical potential splitting, spin-dependent injection current, and applied voltage that can be used to simulate this so-called ``3T" signal as a function of temperature, doping, ferromagnet bulk spin polarization, tunnel barrier features and conduction nonlinearity, and junction voltage bias.
\end{abstract}
\maketitle


\emph{Introduction--} Over the past decade or so, substantial progress has been made in understanding the conditions required for achieving spin-polarized electron transport in otherwise nonmagnetic semiconductors.\cite{Fabian_APS2007} However, the specific constraints imposed\cite{Johnson_PRB1987, Rashba_PRB2000, Schmidt_PRB2000, Slonczewski_PRB1989} empirically necessitate the use of unconventional fabrication techniques\cite{Appelbaum_Nature2007, Huang_PRL2007}, precise control over material growth, and often elaborate high-resolution lithographic and deposition procedures\cite{Lou_NatPhys2007,Suzuki_APE2011}. When it was suggested\cite{Lou_PRL2006} that these complications could be circumvented simply by analyzing the magnetic-field dependence of local magnetoresistance due to spin precession\cite{Hanle_ZPhys1924, Johnson_PRL1985} in large-area single ferromagnet (FM)-insulator-semiconductor tunnel junctions, interest in the experimental spintronics research community was noticeably raised. A significant report in 2009 asserted\cite{Dash_Nature2009} that evidence for spin ``accumulation" in bulk Si persisted through room-temperature in this type of ``3T" device (so-named because of 3-terminal configurations intended to eliminate series ohmic voltage drops through the semiconductor electrode). As a result of these fantastic claims, many in the field quickly and uncritically accepted the new approach as a genuine breakthrough allowing easy access to spin transport properties such as lifetime and diffusion coefficient. \cite{Li_NatureComm2011,Jansen_PRB2010,Jeon_APL2011,Gray_APL2011, Dankert_SciRep2013}

Others, however, were more cautious and exposed inconsistencies that became evident after a more skeptical analysis of the relevant experimental parameters. For instance, spin lifetimes were found to be largely insensitive to doping polarity or concentration, temperature, tunnel barrier material\cite{vantErve_JAP2013}, or semiconductor\cite{Birkner_PRB2013}, all contrary to expectations set by electron spin resonance measurements. Even when normal metals replaced the semiconductor, devices produced the same signals -- independent of the spin-orbit interaction strength which drives relaxation rates!\cite{Txoperena_APL2013} Furthermore, voltage signals are often several orders of magnitude larger than what is possible if due to injection-driven spin accumulation in the bulk. It now appears clear that instead of spin accumulation, it is rather the inelastic transport pathways provided by defect-localized electronic states in or near the tunnel barrier that can play a dominant role in device behavior.\cite{Tran_PRL2009, Jain_PRL2012, Jansen_PRB2012, Tinkey_APL2014, Song_arxiv2014}  

Although the original report by the Minnesota group\cite{Lou_PRL2006} on devices using epitaxial Fe/GaAs, and recent work from the Kyushu group on clean CoFe/Si\cite{Ando_APL2011, Ando_PRB2012, Ishikawa_APL2012, Hamaya_JAP2013} and Kyoto group\cite{Kameno_APL2014} have characteristics (smaller magnitude and dephasing fields, strong temperature dependence, etc.) consistent with true spin accumulation, many specious conclusions on spin transport properties of several important materials have been drawn by others due to a fundamental misunderstanding of the underlying physics behind the 3T technique.  The present paper aims at constructing a rigorous scheme to model such expected genuine results and provide a quantitative means to compare experiment to theory. In particular, we incorporate the essential self-consistency between electrochemical potential splitting, the spin-dependent injection current that induces it, and applied external voltage to simulate magnetoresistance measurements of FM-insulator-semiconductor tunnel junctions as a function of all extrinsic parameters e.g. temperature, doping, ferromagnet bulk spin polarization, tunnel barrier features and conduction nonlinearity, and junction voltage bias.


\emph{Background--} The  steady-state solution to the coupled transport equations for up/down spin electron density $n_{\uparrow/\downarrow}$ for boundary conditions corresponding to spin injection at $z=0$ into a homogeneous semi-infinite conductor $z\ge 0$ with fixed diffusion coefficient $D$, drift velocity $v$, and spin flip rate $1/2\tau$ is $n_\uparrow-n_\downarrow=Pne^{-z/L}$, where $P$ is the electron spin polarization at the injection site, $n=n_\uparrow+n_\downarrow$, and $L=\frac{v\tau}{2}+\sqrt{\left(\frac{v\tau}{2}\right)^2+D\tau}$ is the ``downstream" drift-diffusion transport lengthscale.\cite{Yu_PRB2002} In steady-state, we must supply to this region enough polarized electrons as are lost to spin flips (relaxation-time approximation), resulting in 

\begin{equation}
\frac{(n_\uparrow-n_\downarrow)L}{\tau}=\frac{\beta J}{q},
\label{EQ_RTA}
\end{equation}

\noindent where $J=qnv$ is injected charge current density, $\beta$ is current spin polarization from the FM, and $q$ is the fundamental electron charge. 


\emph{Asymptotic forms--} We will now attempt to estimate the consequences of Eq. \ref{EQ_RTA} for the case of injection into a semiconductor of arbitrary $n$, by making use of simple asymptotic forms of maximum density imbalance $n_\uparrow-n_\downarrow$ in the statistically degenerate and nondegenerate regime. For the former case, thermal energy $k_BT\ll\mu-E_C$, where $\mu$ is chemical potential and $E_C$ is conduction band minimum. Then we have $n_\uparrow-n_\downarrow\simeq \frac{\Delta\mu}{2}\cdot D(\mu-E_C) $, where the density of states at the chemical potential is $D(\mu-E_C)\approx \frac{3n}{2(\mu-E_C)}$, allowing us to calculate the chemical potential splitting

\begin{equation}
\Delta\mu=\mu_\uparrow-\mu_\downarrow\simeq\frac{2(\mu-E_C)}{3} \left[\frac{2\beta J\tau}{qnL}\right]. {\text{   (degenerate)}}
\label{EQ_D}
\end{equation}

For the latter, nondegenerate regime when $k_BT\gg\mu-E_C$, we can use the classical Boltzmann distribution and effective conduction band electron density $N_C=2^{-1/2}(\frac{m_{SC}^*k_BT}{\pi\hbar^2})^{3/2}$ in $n_{\uparrow/\downarrow}=\frac{n}{2}\pm \frac{\beta J\tau}{2qL}\approx\frac{N_C}{2}e^{(\mu_{\uparrow/\downarrow}-E_C)/k_BT}$, giving 

\begin{align}
\mu_{\uparrow/\downarrow}-E_C&\approx k_BT \ln{\left(\frac{n}{N_C}\pm \frac{\beta J\tau}{qLN_C}\right)} \nonumber\\
&=k_BT \left[\ln{\left(\frac{n}{N_C}\right)} + \ln{\left(1\pm\frac{\beta J\tau}{qLn}\right)}\right].\nonumber
\end{align}

For dilute spin densities from weak spin injection or strong relaxation, we can expand the second term to first order, yielding

\begin{equation}
\Delta\mu\simeq k_BT\left[\frac{2\beta J\tau}{qLn}\right]. \qquad {\text{(nondegenerate)}}
\label{EQ_ND}
\end{equation}

\noindent Notice that Eqs. \ref{EQ_D} and \ref{EQ_ND} nominally differ only by the energy scale prefactor. However, nondegenerate systems at low temperatures typically have much smaller density $n$, so that (all other parameters being equal) the unitless quantity in square brackets is much larger in magnitude. Nevertheless, since this quantity is simply the ratio of steady-state spin imbalance to the equilibrium electron density, it is bounded by unity in the regime where the relaxation-time approximation is valid.

This chemical potential splitting $\Delta\mu$ resulting from spin injection is not measured directly in an experiment. In the most na\"ive approach, we can treat injection and detection as separate events. In open-circuit detection appropriate for four-terminal non-local devices with ferromagnetic contacts\cite{Johnson_PRL1985}, injection creates a spin splitting $\Delta\mu$, and due to FM conductance asymmetry (bulk spin polarization) $\beta=\frac{\sigma_\uparrow-\sigma_\downarrow}{\sigma_\uparrow+\sigma_\downarrow}$ a voltage

\begin{equation}
\Delta V=\beta \frac{\Delta \mu}{2 q}
\label{EQ_DV}
\end{equation}

\noindent develops to maintain zero net carrier flow across the interface. Dephasing the spins via precession in a perpendicular field will suppress this voltage, allowing for a direct experimental measurement of $\Delta V$, the additional voltage necessary to drive a fixed current due to spin accumulation magnetoresistance. 

Using Eqs. \ref{EQ_D}, \ref{EQ_ND}, and \ref{EQ_DV}, we can estimate voltage signals of order $\Delta V \apprle \frac{\beta^2 J\tau} {q^2Ln}\mathfrak{E}$, where the energy scale $\mathfrak{E}$ is given by the maximum of the thermal or Fermi energy. While useful to provide gross predictions of an upper bound on expected signals, this approach is however not rigorously correct since we have in fact a biased junction used for \emph{both} injection and detection that cannot be considered separate processes. The spin-dependent chemical potential imbalance necessary for the spin accumulation itself changes the injection rates of spin up and down nonlinearly, so a self-consistent solution is needed. Importantly, this method, freed from reliance on asymptotic expressions used to obtain Eqs. \ref{EQ_D} and \ref{EQ_ND}, will be inherently able to address both \textit{i.}) the intermediate regime between degenerate and nondegenerate conditions where most experiments were performed; and \textit{ii.}) conditions of high spin injection rates, long spin lifetimes, or the deep nondegenerate regime where $\frac{\beta J\tau}{qLn} \sim 1$ (unobtainable in metals) and first-order expansion of the logarithm function used to obtain Eq. \ref{EQ_ND} fails.


\begin{figure}
\begin{center}
\begin{tikzpicture}[scale=1.7]
\node at (-2,3) {(a)};
\draw[black, ultra thick, ->](-1.75,1.75)--(-1.75,3) node[rotate=90, pos=0,left] {Energy};
\draw[black, ultra thick](-1.75,0)--(-1.75,1);

\draw[black, ultra thick](-1.2247,1.5) parabola bend (0,0) (0,0) ;
\draw[black, thick, dashed] (0,0)--(0.25,0) node[right] {$0$};
\draw[black, ultra thick](1,1) parabola bend (1,1) (2,2);
\draw[black, thick](0,0)--(0,3); 
\draw[black, thick,|-|](-0.2,1.5)--(-0.2,3) node[pos=0.5, left] {$\Phi$}; 
\draw[black, thick](1,0)--(1,2); 
\draw[black, thick](0,3)--(1,2); 

\draw[black, thick, dashed](-1.4142,1.5)--(0.4,1.5) node[right] {$E_F$}; 
\draw[black, thick, |-|](0.4,1.5)--(0.4,0.5) node[pos=0.5, left] {$qV$}; 
\draw[black, thick, dashed](1.5,0.5)--(0.4,0.5) node[pos=1.05, left] {$\bar{\mu}$}; 

\draw[black, thick, dashed](1.25,0.8)--(1.75,0.8) node[right] {$\mu_\uparrow$}; 
\draw[black, thick, dashed](1.25,0.2)--(1.75,0.2) node[right] {$\mu_\downarrow$}; 
\draw[black, thick, |-|](1.75,0.2)--(1.75,0.8) node[pos=0.5, right] {$\Delta\mu$}; 
\draw[black, thick, dashed](1,0.65)--(1.3,0.65) node[right] {$\mu_0$}; 

\draw[black, thick, dashed](1,1)--(1.35,1) node[right] {$E_C$}; 

\draw[blue,ultra thick,->](1.15,0.65)--(1.3,0.8);
\draw[blue,ultra thick,->](1.15,0.65)--(1.3,0.2);

\node at (-1,3.05){\uline{FM}};
\node at (1.75,3.05){\uline{SC}};
\node at (0.5,3.05){\uline{insulator}};

\node at (-0.5,0.75){$\frac{\hbar^2k^2}{2m^*_{FM}}$};
\node at (1.5,1.75){$\frac{\hbar^2k^2}{2m^*_{SC}}$};

\draw[black, thick, |-|](0,-0.2)--(1,-0.2) node[pos=0.5, below] {$d$}; 

\draw[black, thick, |-|](0.95,0.5)--(0.95,0.65) node[pos=0.85, left] {$q\Delta V$}; 

\draw[vecArrow](-0.2,1.25) to (1.45,1.25);
\node[red] at (-0.62,1.23){transport};

\node at (-2,-0.3) {(b)};

\end{tikzpicture}

\begin{tikzpicture}[node distance = 1.3cm,auto]
  \node [block] (init) {Initialize P=0, fix $J_0$};
    \node [block, below of=init] (mu) {Find $\mu_{\uparrow/\downarrow}$ [Eq. \ref{EQ_muupdn}]};
    \node [block, right of=mu, node distance=2.5cm] (V) {Find $V$ [Eqs. \ref{EQ_CURR}-\ref{EQ_kS}]};
    \node [block, right of=V, node distance=2.5cm] (J) {Find $J_{\uparrow/\downarrow}$ [Eq. \ref{EQ_RTA}]};
    \node [decision, below of=J, node distance=2cm] (decide) {$\Delta\mu$ \nohyphens{converged?}};
    \node [block, below of=decide,node distance=2cm] (stop) {Calc. $\Delta V$ [Eq. \ref{EQ_CALCDV}]};
    \path [line] (init) -- (mu);
    \path [line] (mu) -- (V);
    \path [line] (V) -- (J);
    \path [line] (J) -- (decide);
    
    
\path [line] (decide) -| node [above,pos=0.05]{no}(mu);

\path [line] (decide) -- node {yes}(stop);
    \end{tikzpicture}
    \vspace{-10pt}
\end{center}
\caption{(Color online) (a) Schematic energy band diagram identifying parameters in the model simulating spin accumulation magnetoresistance in the ferromagnet (FM)-insulator-semiconductor (SC) tunnel junction. Blue arrows show equilibrium electrochemical potential $\mu_0$ splitting due to spin-polarized electron injection. (b) Schematic flow diagram for the self-consistent algorithm used to calculate ``3T" magnetoresistance signal $\Delta V$ at constant current density $J_0$ and electron density $n$ due to spin accumulation and precession in a magnetic field.\label{FIG_BAND}}
\vspace{-10pt}
\end{figure}
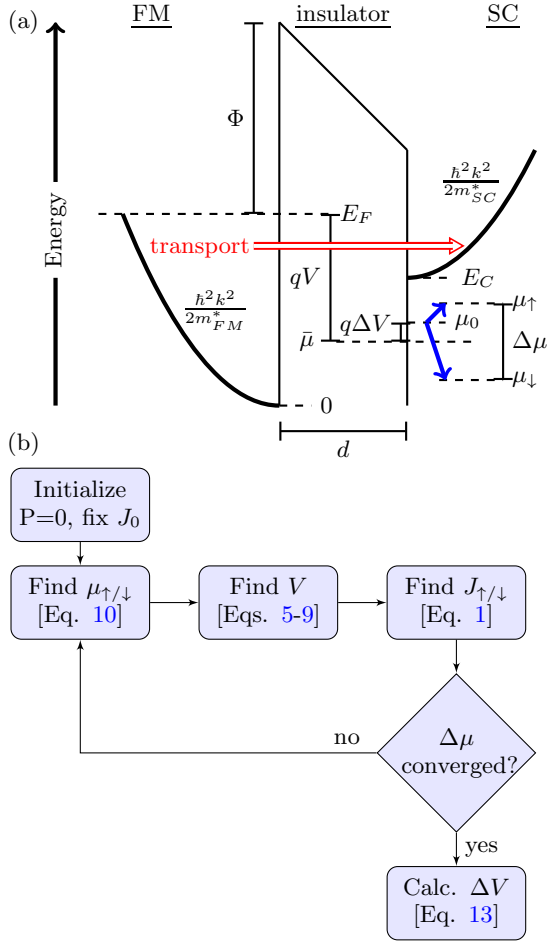

\emph{Self-consistent model--} Consider the FM-semiconductor tunnel junction illustrated in Fig. \ref{FIG_BAND}. Since our focus here is on a purely elastic tunneling model, we ignore Fermi level pinning from interface states resulting in thermionic emission-dominated transport into the depletion region, which would otherwise potentially complicate detection \cite{Jansen_PRL2007}.

To begin the self-consistent calculation, we initialize the unknown $P=\frac{n_\uparrow-n_\downarrow}{n_\uparrow+n_\downarrow}=0$ so that $n_\uparrow=n_\downarrow=n/2$ and $\mu_\uparrow=\mu_\downarrow=\mu_0$ in the conduction band. We can then calculate the current density $J=J_0=J_\uparrow+J_\downarrow$ which flows under voltage $V=V_0$, using an extension of the 1-dimensional transport model\cite{Harrison_PR1961} which sums over FM cathode states:

\begin{align} 
J_{\uparrow/\downarrow}=&q\frac{1\pm\beta}{2} \frac{2m_{FM}^*E_F}{h^3}\times \nonumber\\
&\int_0^{\infty}\left[f_T(E-E_F)-f_T(E-\mu_{\uparrow/\downarrow})\right]\mathfrak{T}(E,V)dE,
\label{EQ_CURR}
\end{align} 

\noindent where $f_T$ is the Fermi-Dirac occupation function at temperature $T$, and we employ the usual semiclassical approximation for incoherent tunneling transmission coefficient $\mathfrak{T}$,\cite{Slonczewski_PRB1989} correct to lowest order in the ratios of $E$  and $qV$ to the total tunnel barrier height $E_F+\Phi$, where $\Phi$ is the tunnel barrier internal work function:

\begin{align} 
\mathfrak{T}(E,V)&=\frac{4k_{FM}\kappa}{k_{FM}^2+\kappa^2}\frac{4k_{SC}\kappa}{k_{SC}^2+\kappa^2}e^{-2\kappa d},{\text{ where}}
\end{align}
\begin{align}
\kappa&=\sqrt{2m_I^*(E_F+\Phi-qV/2-E)}/\hbar,\\
k_{FM}&=\sqrt{2m_{FM}^*E}/\hbar, {\text{and}}\\
k_{SC}&=\Re{\sqrt{2m_{SC}^*(E-E_C)}/\hbar}. \label{EQ_kS}
\end{align}

\noindent Notice that $k_{SC}$, and therefore $\mathfrak{T}$, vanishes for electrons with energy in the forbidden gap.

Now, we perform the following four steps in a loop to obtain self-consistency between the injection rate $J$ and the spin accumulation $\Delta\mu$ by iterative adjustment of the junction bias voltage $V$ (see Fig. \ref{FIG_BAND} (b)):

\emph{[1.]} Using the present value of accumulated polarization $P$, find $\mu_\uparrow$ and $\mu_\downarrow$ consistent with $n=n_\uparrow+n_\downarrow$ from the sum over semiconductor states 
\begin{align}
&n_{\uparrow/\downarrow}=\notag\\
&\frac{(1\pm P)n}{2}=g\int_0^\infty{f_T(E-\mu_{\uparrow/\downarrow})D_{\frac{1}{2}}(E-E_C)dE}, \label{EQ_muupdn}
\end{align}
\noindent where $g$ is the conduction band degeneracy (e.g. $g=6$ for Si, $=4$ for Ge), and the single-spin density of states $D_{\frac{1}{2}}(E)=\frac{1}{\sqrt{2}\pi^2}\left(\frac{m^*_{SC}}{\hbar^2}\right)^{3/2} \sqrt{E}$. This nonlinear inverse problem is solved using a binomial search algorithm.

\emph{[2.]}
Using the new values for $\mu_{\uparrow/\downarrow}$, employ Eqs. \ref{EQ_CURR}-\ref{EQ_kS} to find the applied voltage $V > V_0$ necessary to maintain fixed current density $J_0=J_\uparrow+J_\downarrow$. Again, this is a nonlinear inverse problem, which can be solved using a binomial search algorithm.

\emph{[3.]}
Using the new values for spin-dependent current densities $J_{\uparrow/\downarrow}$, update the accumulated polarization from the relaxation-time approximation result (Eq. \ref{EQ_RTA}) in the form $P=\frac{(J_\uparrow-J_\downarrow) \tau}{qnL}$. By rewriting the drift-diffusion lengthscale as

\begin{equation}
L=\sqrt{D\tau+\left(\frac{J_0\tau}{2nq}\right)^2} +\frac{J_0\tau}{2nq},
\label{Eq_Ld}
\end{equation}

\noindent we see that in the high injection current limit, even if $\beta=1$, this implies (to lowest order) 

\begin{equation}
P\apprle 1-\frac{q^2n^2D}{\tau J_0^2}.
\label{Eq_P}
\end{equation}

The spin polarization is thus always strictly less than unity as required.

\emph{[4.]}
Repeat steps 1-3 until convergence of $\mu_{\uparrow/\downarrow}$. When completed, the self-consistent voltage necessary in constant-current mode is the difference between the spin-independent (average) chemical potential in the SC, $\bar\mu= \frac{\mu_\uparrow+ \mu_\downarrow}{2}$, and the FM Fermi energy, implying a change in applied voltage in a perpendicular magnetic field necessary to fully dephase the spins of 

\begin{equation}
\Delta V=\frac{E_F-\bar\mu}{q}-V_0.
\label{EQ_CALCDV}
\end{equation}

Given the structure of expressions above, we qualitatively expect that self consistent iteration is most important under a particular set of circumstances. Of course, this includes the condition that $\Delta V$ is comparable to $V$, so that changes in $\Delta \mu$ affect large changes in the ratios of spin up/down injected current. Usually, this requires highly conductive tunnel barriers and large injected current spin polarization $\beta$. However, Eq. \ref{Eq_P} shows that the steady-state density polarization $P$ saturates at high $J_0$, making self-consistency unnecessary beyond one iteration in this limit. Therefore, assuming a fixed $n$, $\tau$, and $D$, a self-consistent approach is most important when $P$ and $\frac{dP}{dJ}$ are jointly maximized, which occurs for 

\begin{equation}
J^*_0\approx qn\sqrt{2D/\tau}.
\label{EQ_OPT}
\end{equation}

\noindent This value is recognizable as the current for which the contributions from drift and diffusion are approximately equal in Eq. \ref{Eq_Ld}.




\begin{figure}
\begin{center}
\includegraphics[scale=0.425]{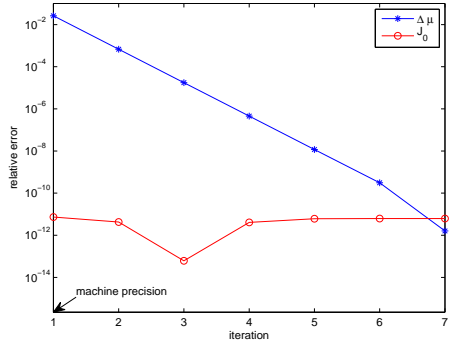}
\vspace{-10pt}
\caption{(Color online) Convergence of the self-consistent scheme, showing the exponential reduction of relative error with iteration of calculation steps 1-4 discussed in the text. \label{FIG_CONV}} 
\vspace{-20pt}
\end{center}
\end{figure}

\emph{Results--} Now we discuss the results of our self-consistent method. In Fig. \ref{FIG_CONV}, we see an initial error relative to converged value of several percent in $\Delta\mu$. However, we also see fast exponential convergence toward negligible error upon successive iteration. Here, we use parameters $E_F=5$eV, $d=1$nm, $\Phi=1$eV,  $n=10^{19}$cm$^{-3}$ (just below ambient $N_C\approx 3\times 10^{19}$), $m^*_{FM}=m_0$, and $m^*_{SC}=0.36m_0$. At $T=300$K, $V_0=100$mV drives a current density $J_0\approx 3\times 10^4$A/cm$^2$, on the order of $J_0^*$ given in Eq. \ref{EQ_OPT}. With a FM polarization of $\beta=0.4$, $D=10$cm$^2$/s, and $\tau=10$ns, we obtain a steady state spin accumulation in the SC of approximately $P=0.17$ and $\Delta\mu\approx 0.6$mV. Note that, since these parameters give $\frac{2\beta J \tau}{qnL}\approx 0.5\sim 1$, the results represent the extreme high-injection limit of validity and are used here merely to illustrate the model behavior at current densities close to $J^*_0$ (Eq. \ref{EQ_OPT}). Also shown in Fig. \ref{FIG_CONV} is the small relative error of the tunnel current density during the voltage adjustment, illustrating the stability of the method and effectiveness of the binomial search algorithm as applied to Eq. \ref{EQ_CURR}.

\begin{figure}
\begin{center}
\includegraphics[scale=0.45]{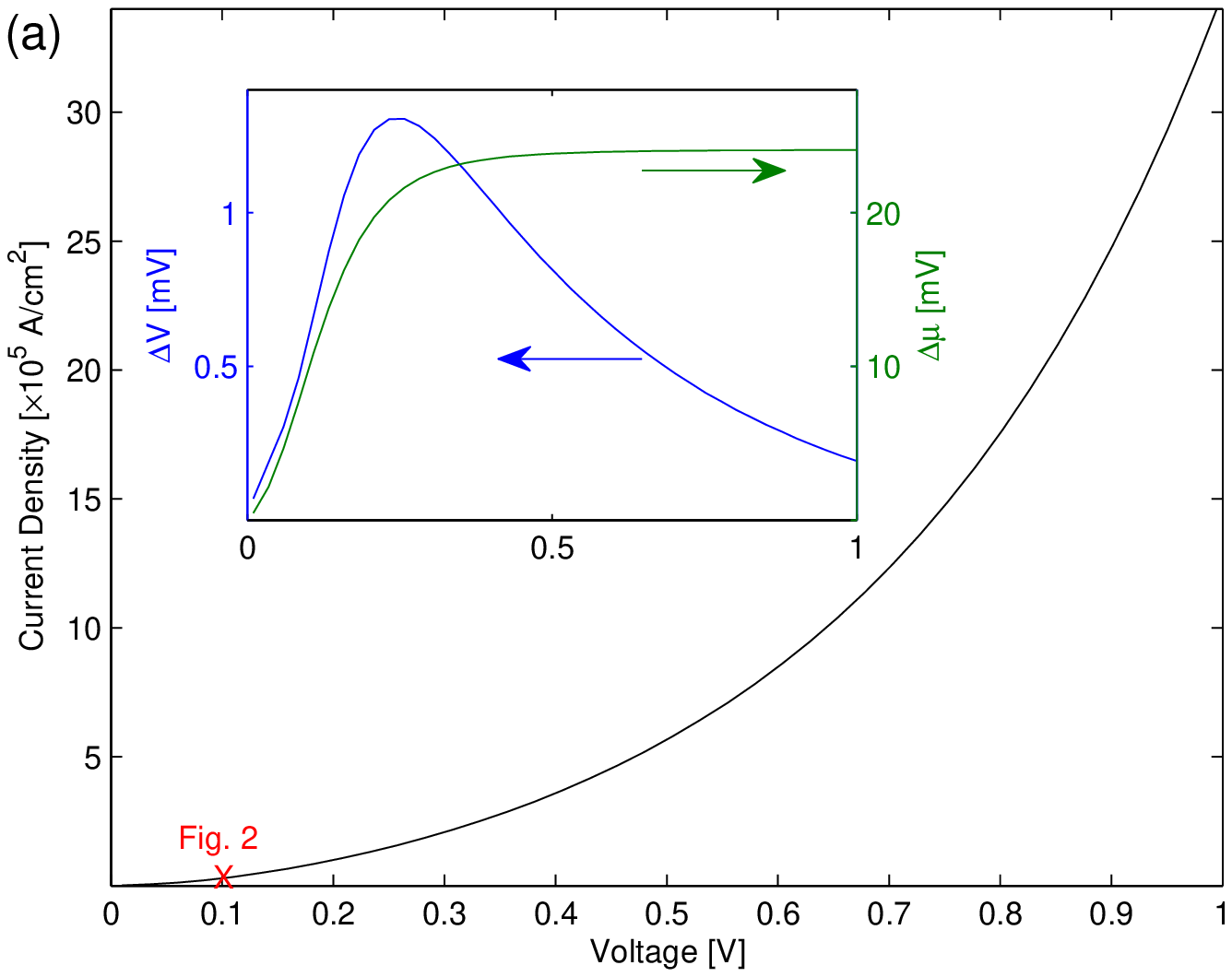}
\includegraphics[scale=0.45]{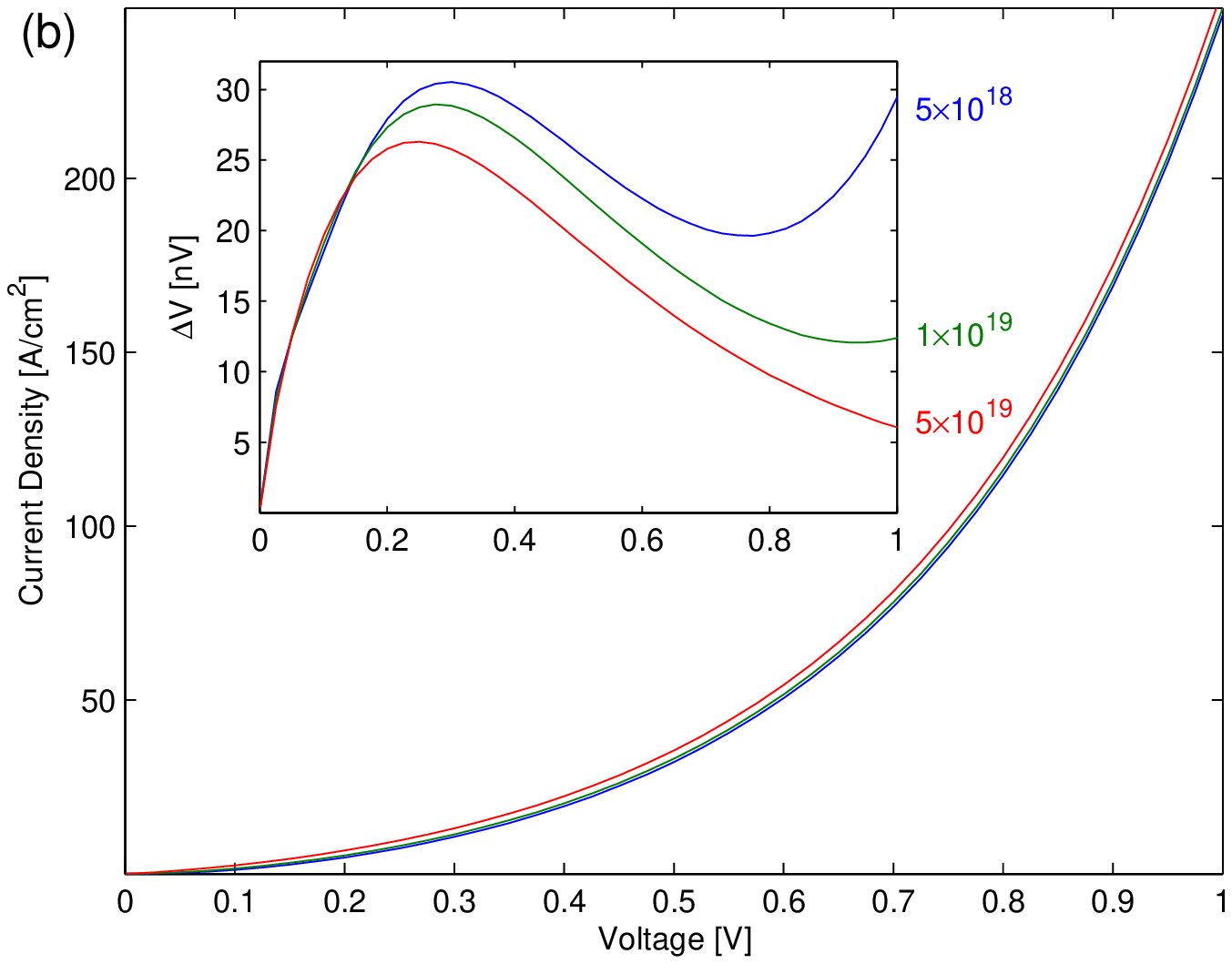}
\vspace{-20pt}
\end{center}
\caption{(Color online) (a) Current-voltage relation in the strong injection current regime. Self-consistent convergence data shown in Fig. 2 for point indicated. Inset: Corresponding spin accumulation voltage and electrochemical splitting. (b) I-V in the more realistic weak injection regime, showing little dependence on donor concentration. Inset: Spin accumulation signals over the same voltage range. \label{FIG_IV}}
\vspace{-15pt}
\end{figure}

The current-voltage relationship for the same junction is shown in Fig. \ref{FIG_IV}(a), where the nonlinear behavior expected from tunneling can clearly be seen. In the inset, the corresponding voltage dependence of converged variables shows that increasing the current injection with higher voltage results in saturation of the accumulated polarization ($\propto \Delta \mu$) and a suppression of spin accumulation signal ($\Delta V$). This latter phenomenon can be understood by considering the effect of an increasing conductance nonlinearity at high voltage on the constraints of fixed current density while reaching self-consistency.


The calculations presented so far have used parameters intended to approach the current density in Eq. \ref{EQ_OPT}. Actual experiments performed to date have, however, far smaller tunnel junction conductances. By adjusting $d=1.5$nm and $\Phi=$1.7eV (appropriate for Al$_2$O$_3$ barriers), the current density dramatically decreases to more realistic values, as shown for several SC donor densities in Fig. \ref{FIG_IV}(b). Although the current-voltage relationship is only weakly dependent on doping in this restricted range, the spin accumulation voltage signals shown in the inset clearly demonstrate a stronger sensitivity. Importantly, we see here that the magnitude of $\Delta V$ (tens of nV) is orders of magnitude smaller than in the high injection limit above; for a comparison to the asymptotic expressions given earlier, Eqs. \ref{EQ_ND} and \ref{EQ_DV} predict $\Delta V\approx 8\mu$V for $n=5\times 10^{18}$cm$^{-3}$, and using Eq. \ref{EQ_D} for $n=5\times 10^{19}$cm$^{-3}$ yields $\Delta V\approx 500$nV. Both degenerate and nondegenerate asymptotic approaches sharply overestimate the expected accumulation signal for these intermediate electron densities.

\begin{figure}
\begin{center}
\includegraphics[scale=0.45]{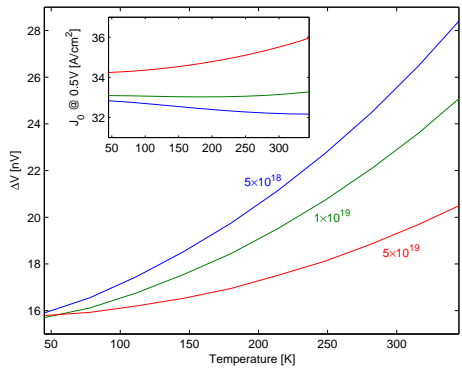}
\vspace{-10pt}
\caption{(Color online) Temperature dependence of spin accumulation voltage signals at a fixed $V_0=$0.5V. Inset: current density at this bias. \label{FIG_temp}} 
\vspace{-20pt}
\end{center}
\end{figure}

Temperature dependence of the spin accumulation voltage signals is shown in Fig. \ref{FIG_temp} with a reference voltage of $V_0=$0.5V. The current density varies only slightly over this range (inset), but $\Delta V$ increases dramatically with temperature over $\approx$100K. This behavior (constant at low temperatures in the quasi-degenerate regime and linearly increasing in the quasi-nondegenerate regime) is consistent with the trends predicted by Eqs. \ref{EQ_D} and \ref{EQ_ND}. However, the calculated signal magnitudes are much smaller than indicated by the coarse approximations.  

\begin{figure}
\begin{center}
\includegraphics[scale=0.45]{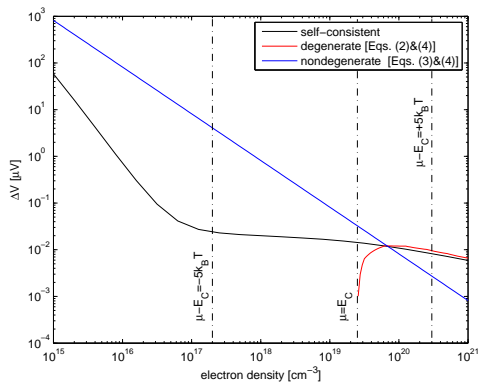}
\caption{(Color online) comparison of asymptotic forms to self-consistent values for ``3T" magnetoresistance voltage $\Delta V$, for $J_0=$1~A/cm$^2$ at 300K. \label{FIG_nvsDV}} 
\vspace{-20pt}
\end{center}
\end{figure}
\vspace{10pt}

\emph{Conclusion--} To illustrate this apparent discrepancy, we compare the asymptotic forms at 300K to the calculated self-consistent results for low injection $J_0=1$A/cm$^2$ as a function of electron density $n$ in Fig. \ref{FIG_nvsDV}. We have highlighted three regions, separated by two vertical lines, where $\mu-E_C<-5k_BT$ (nondegenerate), $\mu-E_C>+5k_BT$ (degenerate), and the intermediate regime where $\mu\approx E_C$. Clearly, the numerically calculated results trend toward the asymptotic curves in the doping level extremes, with especially good approximation in the degenerate limit. However, the intermediate and weakly nondegenerate regime show two orders of magnitude of disagreement. Even at the solidly-nondegenerate density of $n=10^{15}$cm$^{-3}$, at least an order of magnitude separates the results. 

The lack of agreement for low electron densities can be understood by considering that as $n$ decreases, junction zero-bias resistance increases and the tunnel bias is substantially larger for the same driven current density $J_0$. The junction nonlinearity is then appreciable, and the suppression of $\Delta V$ seen at high bias in the insets to Figs. \ref{FIG_IV}(a) and (b) becomes substantial. This explanation is corroborated by the fact that, all other parameters being equal, deviation from Eqs. \ref{EQ_D} and \ref{EQ_ND} across all densities worsens as $J_0$ (and hence $V_0$) is chosen to be larger. The problem at lower electron doping makes it clear that the asymptotic expression for the nondegenerate case Eq. \ref{EQ_ND} is merely an optimistic upper bound, and that realistic signals from devices in this regime (more correctly simulated with the self-consistent scheme described here) will in fact be systematically much lower. Even for the degenerate case, voltage changes from spin-accumulation magnetoresistance induced by precession and dephasing are far lower than those experimental measurements attributed to this mechanism. For example, the original claim of room-temperature spin accumulation in 1.8$\times 10^{19}$cm$^{-3}$ n-Si with $\Delta V\approx 1.2$mV for $J=4$A/cm$^2$ (see Ref. \onlinecite{Dash_Nature2009}) is so far above physically reasonable values that it (and the results of other room-temperature measurements\cite{Li_NatureComm2011,Jansen_PRB2010,Jeon_APL2011,Gray_APL2011, Dankert_SciRep2013}) is vertically off-scale in Fig. \ref{FIG_nvsDV}.

We gratefully acknowledge support from ONR under contract N000141410317, NSF under contracts  ECCS-1231855 and DGE-1322106 (H.N.T.), and DTRA under contract HDTRA1-13-1-0013. 


\begin{thebibliography}{34}%
\makeatletter
\providecommand \@ifxundefined [1]{%
 \@ifx{#1\undefined}
}%
\providecommand \@ifnum [1]{%
 \ifnum #1\expandafter \@firstoftwo
 \else \expandafter \@secondoftwo
 \fi
}%
\providecommand \@ifx [1]{%
 \ifx #1\expandafter \@firstoftwo
 \else \expandafter \@secondoftwo
 \fi
}%
\providecommand \natexlab [1]{#1}%
\providecommand \enquote  [1]{``#1''}%
\providecommand \bibnamefont  [1]{#1}%
\providecommand \bibfnamefont [1]{#1}%
\providecommand \citenamefont [1]{#1}%
\providecommand \href@noop [0]{\@secondoftwo}%
\providecommand \href [0]{\begingroup \@sanitize@url \@href}%
\providecommand \@href[1]{\@@startlink{#1}\@@href}%
\providecommand \@@href[1]{\endgroup#1\@@endlink}%
\providecommand \@sanitize@url [0]{\catcode `\\12\catcode `\$12\catcode
  `\&12\catcode `\#12\catcode `\^12\catcode `\_12\catcode `\%12\relax}%
\providecommand \@@startlink[1]{}%
\providecommand \@@endlink[0]{}%
\providecommand \url  [0]{\begingroup\@sanitize@url \@url }%
\providecommand \@url [1]{\endgroup\@href {#1}{\urlprefix }}%
\providecommand \urlprefix  [0]{URL }%
\providecommand \Eprint [0]{\href }%
\providecommand \doibase [0]{http://dx.doi.org/}%
\providecommand \selectlanguage [0]{\@gobble}%
\providecommand \bibinfo  [0]{\@secondoftwo}%
\providecommand \bibfield  [0]{\@secondoftwo}%
\providecommand \translation [1]{[#1]}%
\providecommand \BibitemOpen [0]{}%
\providecommand \bibitemStop [0]{}%
\providecommand \bibitemNoStop [0]{.\EOS\space}%
\providecommand \EOS [0]{\spacefactor3000\relax}%
\providecommand \BibitemShut  [1]{\csname bibitem#1\endcsname}%
\let\auto@bib@innerbib\@empty
\bibitem [{\citenamefont {Fabian}\ \emph {et~al.}(2007)\citenamefont {Fabian}
  \emph {et~al.}}]{Fabian_APS2007}%
  \BibitemOpen
  \bibfield  {author} {\bibinfo {author} {\bibfnamefont {J.}~\bibnamefont
  {Fabian}} \emph {et~al.},\ }\href {\doibase 10.2478/v10155-010-0086-8}
  {\bibfield  {journal} {\bibinfo  {journal} {Acta Phys. Slovaca}\ }\textbf
  {\bibinfo {volume} {57}},\ \bibinfo {pages} {565} (\bibinfo {year}
  {2007})}\BibitemShut {NoStop}%
\bibitem [{\citenamefont {Johnson}\ and\ \citenamefont
  {Silsbee}(1987)}]{Johnson_PRB1987}%
  \BibitemOpen
  \bibfield  {author} {\bibinfo {author} {\bibfnamefont {M.}~\bibnamefont
  {Johnson}}\ and\ \bibinfo {author} {\bibfnamefont {R.}~\bibnamefont
  {Silsbee}},\ }\href {\doibase 10.1103/PhysRevB.35.4959} {\bibfield  {journal}
  {\bibinfo  {journal} {Phys. Rev. B}\ }\textbf {\bibinfo {volume} {35}},\
  \bibinfo {pages} {4959} (\bibinfo {year} {1987})}\BibitemShut {NoStop}%
\bibitem [{\citenamefont {Rashba}(2000)}]{Rashba_PRB2000}%
  \BibitemOpen
  \bibfield  {author} {\bibinfo {author} {\bibfnamefont {E.}~\bibnamefont
  {Rashba}},\ }\href {\doibase 10.1103/PhysRevB.62.R16267} {\bibfield
  {journal} {\bibinfo  {journal} {Phys. Rev. B}\ }\textbf {\bibinfo {volume}
  {62}},\ \bibinfo {pages} {R16267} (\bibinfo {year} {2000})}\BibitemShut
  {NoStop}%
\bibitem [{\citenamefont {Schmidt}\ \emph {et~al.}(2000)\citenamefont {Schmidt}
  \emph {et~al.}}]{Schmidt_PRB2000}%
  \BibitemOpen
  \bibfield  {author} {\bibinfo {author} {\bibfnamefont {G.}~\bibnamefont
  {Schmidt}} \emph {et~al.},\ }\href {\doibase 10.1103/PhysRevB.62.R4790}
  {\bibfield  {journal} {\bibinfo  {journal} {Phys. Rev. B}\ }\textbf {\bibinfo
  {volume} {62}},\ \bibinfo {pages} {R4790} (\bibinfo {year}
  {2000})}\BibitemShut {NoStop}%
\bibitem [{\citenamefont {Slonczewski}(1989)}]{Slonczewski_PRB1989}%
  \BibitemOpen
  \bibfield  {author} {\bibinfo {author} {\bibfnamefont {J.~C.}\ \bibnamefont
  {Slonczewski}},\ }\href {\doibase 10.1103/PhysRevB.39.6995} {\bibfield
  {journal} {\bibinfo  {journal} {Phys. Rev. B}\ }\textbf {\bibinfo {volume}
  {39}},\ \bibinfo {pages} {6995} (\bibinfo {year} {1989})}\BibitemShut
  {NoStop}%
\bibitem [{\citenamefont {Appelbaum}\ \emph {et~al.}(2007)\citenamefont
  {Appelbaum} \emph {et~al.}}]{Appelbaum_Nature2007}%
  \BibitemOpen
  \bibfield  {author} {\bibinfo {author} {\bibfnamefont {I.}~\bibnamefont
  {Appelbaum}} \emph {et~al.},\ }\href {\doibase 10.1038/nature05803}
  {\bibfield  {journal} {\bibinfo  {journal} {Nature}\ }\textbf {\bibinfo
  {volume} {447}},\ \bibinfo {pages} {295} (\bibinfo {year}
  {2007})}\BibitemShut {NoStop}%
\bibitem [{\citenamefont {Huang}\ \emph {et~al.}(2007)\citenamefont {Huang}
  \emph {et~al.}}]{Huang_PRL2007}%
  \BibitemOpen
  \bibfield  {author} {\bibinfo {author} {\bibfnamefont {B.}~\bibnamefont
  {Huang}} \emph {et~al.},\ }\href {\doibase 10.1103/PhysRevLett.99.177209}
  {\bibfield  {journal} {\bibinfo  {journal} {Phys. Rev. Lett.}\ }\textbf
  {\bibinfo {volume} {99}},\ \bibinfo {pages} {177209} (\bibinfo {year}
  {2007})}\BibitemShut {NoStop}%
\bibitem [{\citenamefont {Lou}\ \emph {et~al.}(2007)\citenamefont {Lou} \emph
  {et~al.}}]{Lou_NatPhys2007}%
  \BibitemOpen
  \bibfield  {author} {\bibinfo {author} {\bibfnamefont {X.}~\bibnamefont
  {Lou}} \emph {et~al.},\ }\href {\doibase 10.1038/nphys543} {\bibfield
  {journal} {\bibinfo  {journal} {Nat. Phys.}\ }\textbf {\bibinfo {volume}
  {3}},\ \bibinfo {pages} {197} (\bibinfo {year} {2007})}\BibitemShut {NoStop}%
\bibitem [{\citenamefont {Suzuki}\ \emph {et~al.}(2011)\citenamefont {Suzuki}
  \emph {et~al.}}]{Suzuki_APE2011}%
  \BibitemOpen
  \bibfield  {author} {\bibinfo {author} {\bibfnamefont {T.}~\bibnamefont
  {Suzuki}} \emph {et~al.},\ }\href {\doibase 10.1143/APEX.4.023003} {\bibfield
   {journal} {\bibinfo  {journal} {Appl. Phys. Express}\ }\textbf {\bibinfo
  {volume} {4}},\ \bibinfo {pages} {3} (\bibinfo {year} {2011})}\BibitemShut
  {NoStop}%
\bibitem [{\citenamefont {Lou}\ \emph {et~al.}(2006)\citenamefont {Lou} \emph
  {et~al.}}]{Lou_PRL2006}%
  \BibitemOpen
  \bibfield  {author} {\bibinfo {author} {\bibfnamefont {X.}~\bibnamefont
  {Lou}} \emph {et~al.},\ }\href {\doibase 10.1103/PhysRevLett.96.176603}
  {\bibfield  {journal} {\bibinfo  {journal} {Phys. Rev. Lett.}\ }\textbf
  {\bibinfo {volume} {96}},\ \bibinfo {pages} {176603} (\bibinfo {year}
  {2006})}\BibitemShut {NoStop}%
\bibitem [{\citenamefont {Hanle}(1924)}]{Hanle_ZPhys1924}%
  \BibitemOpen
  \bibfield  {author} {\bibinfo {author} {\bibfnamefont {W.}~\bibnamefont
  {Hanle}},\ }\href {\doibase 10.1007/BF01331827} {\bibfield  {journal}
  {\bibinfo  {journal} {Z. Phys.}\ }\textbf {\bibinfo {volume} {30}},\ \bibinfo
  {pages} {93} (\bibinfo {year} {1924})}\BibitemShut {NoStop}%
\bibitem [{\citenamefont {Johnson}\ and\ \citenamefont
  {Silsbee}(1985)}]{Johnson_PRL1985}%
  \BibitemOpen
  \bibfield  {author} {\bibinfo {author} {\bibfnamefont {M.}~\bibnamefont
  {Johnson}}\ and\ \bibinfo {author} {\bibfnamefont {R.}~\bibnamefont
  {Silsbee}},\ }\href {\doibase 10.1103/PhysRevLett.55.1790} {\bibfield
  {journal} {\bibinfo  {journal} {Phys. Rev. Lett.}\ }\textbf {\bibinfo
  {volume} {55}},\ \bibinfo {pages} {1790} (\bibinfo {year}
  {1985})}\BibitemShut {NoStop}%
\bibitem [{\citenamefont {Dash}\ \emph {et~al.}(2009)\citenamefont {Dash} \emph
  {et~al.}}]{Dash_Nature2009}%
  \BibitemOpen
  \bibfield  {author} {\bibinfo {author} {\bibfnamefont {S.~P.}\ \bibnamefont
  {Dash}} \emph {et~al.},\ }\href {\doibase 10.1038/nature08570} {\bibfield
  {journal} {\bibinfo  {journal} {Nature}\ }\textbf {\bibinfo {volume} {462}},\
  \bibinfo {pages} {245} (\bibinfo {year} {2009})}\BibitemShut {NoStop}%
\bibitem [{\citenamefont {Li}\ \emph {et~al.}(2011)\citenamefont {Li} \emph
  {et~al.}}]{Li_NatureComm2011}%
  \BibitemOpen
  \bibfield  {author} {\bibinfo {author} {\bibfnamefont {C.}~\bibnamefont {Li}}
  \emph {et~al.},\ }\href {\doibase 10.1038/ncomms1256} {\bibfield  {journal}
  {\bibinfo  {journal} {Nature Comm.}\ }\textbf {\bibinfo {volume} {2}},\
  \bibinfo {pages} {245} (\bibinfo {year} {2011})}\BibitemShut {NoStop}%
\bibitem [{\citenamefont {Jansen}\ \emph {et~al.}(2010)\citenamefont {Jansen}
  \emph {et~al.}}]{Jansen_PRB2010}%
  \BibitemOpen
  \bibfield  {author} {\bibinfo {author} {\bibfnamefont {R.}~\bibnamefont
  {Jansen}} \emph {et~al.},\ }\href {\doibase 10.1103/PhysRevB.82.241305}
  {\bibfield  {journal} {\bibinfo  {journal} {Phys. Rev. B}\ }\textbf {\bibinfo
  {volume} {82}},\ \bibinfo {pages} {241305(R)} (\bibinfo {year}
  {2010})}\BibitemShut {NoStop}%
\bibitem [{\citenamefont {Jeon}\ \emph {et~al.}(2011)\citenamefont {Jeon} \emph
  {et~al.}}]{Jeon_APL2011}%
  \BibitemOpen
  \bibfield  {author} {\bibinfo {author} {\bibfnamefont {K.}~\bibnamefont
  {Jeon}} \emph {et~al.},\ }\href {\doibase 10.1063/1.3600787} {\bibfield
  {journal} {\bibinfo  {journal} {Appl. Phys. Lett.}\ }\textbf {\bibinfo
  {volume} {98}},\ \bibinfo {pages} {3} (\bibinfo {year} {2011})}\BibitemShut
  {NoStop}%
\bibitem [{\citenamefont {Gray}\ and\ \citenamefont
  {Tiwari}(2011)}]{Gray_APL2011}%
  \BibitemOpen
  \bibfield  {author} {\bibinfo {author} {\bibfnamefont {N.}~\bibnamefont
  {Gray}}\ and\ \bibinfo {author} {\bibfnamefont {A.}~\bibnamefont {Tiwari}},\
  }\href {\doibase 10.1063/1.3564889} {\bibfield  {journal} {\bibinfo
  {journal} {Appl. Phys. Lett.}\ }\textbf {\bibinfo {volume} {98}},\ \bibinfo
  {pages} {102112} (\bibinfo {year} {2011})}\BibitemShut {NoStop}%
\bibitem [{\citenamefont {Dankert}\ \emph {et~al.}(2013)\citenamefont {Dankert}
  \emph {et~al.}}]{Dankert_SciRep2013}%
  \BibitemOpen
  \bibfield  {author} {\bibinfo {author} {\bibfnamefont {A.}~\bibnamefont
  {Dankert}} \emph {et~al.},\ }\href {\doibase 10.1038/srep03196} {\bibfield
  {journal} {\bibinfo  {journal} {Sci. Rep.}\ }\textbf {\bibinfo {volume}
  {3}},\ \bibinfo {pages} {3196} (\bibinfo {year} {2013})}\BibitemShut
  {NoStop}%
\bibitem [{\citenamefont {van't Erve}\ \emph {et~al.}(2013)\citenamefont {van't
  Erve} \emph {et~al.}}]{vantErve_JAP2013}%
  \BibitemOpen
  \bibfield  {author} {\bibinfo {author} {\bibfnamefont {O.}~\bibnamefont
  {van't Erve}} \emph {et~al.},\ }\href {\doibase
  http://dx.doi.org/10.1063/1.4793712} {\bibfield  {journal} {\bibinfo
  {journal} {J. Appl. Phys.}\ }\textbf {\bibinfo {volume} {113}},\ \bibinfo
  {eid} {17C502} (\bibinfo {year} {2013})}\BibitemShut {NoStop}%
\bibitem [{\citenamefont {Birkner}\ \emph {et~al.}(2013)\citenamefont {Birkner}
  \emph {et~al.}}]{Birkner_PRB2013}%
  \BibitemOpen
  \bibfield  {author} {\bibinfo {author} {\bibfnamefont {B.}~\bibnamefont
  {Birkner}} \emph {et~al.},\ }\href {\doibase 10.1103/PhysRevB.87.081405}
  {\bibfield  {journal} {\bibinfo  {journal} {Phys. Rev. B}\ }\textbf {\bibinfo
  {volume} {87}},\ \bibinfo {pages} {081405} (\bibinfo {year}
  {2013})}\BibitemShut {NoStop}%
\bibitem [{\citenamefont {Txoperena}\ \emph {et~al.}(2013)\citenamefont
  {Txoperena} \emph {et~al.}}]{Txoperena_APL2013}%
  \BibitemOpen
  \bibfield  {author} {\bibinfo {author} {\bibfnamefont {O.}~\bibnamefont
  {Txoperena}} \emph {et~al.},\ }\href {\doibase 10.1063/1.4806987} {\bibfield
  {journal} {\bibinfo  {journal} {Appl. Phys. Lett.}\ }\textbf {\bibinfo
  {volume} {102}},\ \bibinfo {pages} {192406} (\bibinfo {year}
  {2013})}\BibitemShut {NoStop}%
\bibitem [{\citenamefont {Tran}\ \emph {et~al.}(2009)\citenamefont {Tran} \emph
  {et~al.}}]{Tran_PRL2009}%
  \BibitemOpen
  \bibfield  {author} {\bibinfo {author} {\bibfnamefont {M.}~\bibnamefont
  {Tran}} \emph {et~al.},\ }\href {\doibase 10.1103/PhysRevLett.102.036601}
  {\bibfield  {journal} {\bibinfo  {journal} {Phys. Rev. Lett.}\ }\textbf
  {\bibinfo {volume} {102}},\ \bibinfo {pages} {036601} (\bibinfo {year}
  {2009})}\BibitemShut {NoStop}%
\bibitem [{\citenamefont {Jain}\ \emph {et~al.}(2012)\citenamefont {Jain} \emph
  {et~al.}}]{Jain_PRL2012}%
  \BibitemOpen
  \bibfield  {author} {\bibinfo {author} {\bibfnamefont {A.}~\bibnamefont
  {Jain}} \emph {et~al.},\ }\href {\doibase 10.1103/PhysRevLett.109.106603}
  {\bibfield  {journal} {\bibinfo  {journal} {Phys. Rev. Lett.}\ }\textbf
  {\bibinfo {volume} {109}},\ \bibinfo {pages} {106603} (\bibinfo {year}
  {2012})}\BibitemShut {NoStop}%
\bibitem [{\citenamefont {Jansen}\ \emph {et~al.}(2012)\citenamefont {Jansen}
  \emph {et~al.}}]{Jansen_PRB2012}%
  \BibitemOpen
  \bibfield  {author} {\bibinfo {author} {\bibfnamefont {R.}~\bibnamefont
  {Jansen}} \emph {et~al.},\ }\href {\doibase 10.1103/PhysRevB.85.134420}
  {\bibfield  {journal} {\bibinfo  {journal} {Phys. Rev. B}\ }\textbf {\bibinfo
  {volume} {85}},\ \bibinfo {pages} {134420} (\bibinfo {year}
  {2012})}\BibitemShut {NoStop}%
\bibitem [{\citenamefont {Tinkey}\ \emph {et~al.}(2014)\citenamefont {Tinkey}
  \emph {et~al.}}]{Tinkey_APL2014}%
  \BibitemOpen
  \bibfield  {author} {\bibinfo {author} {\bibfnamefont {H.}~\bibnamefont
  {Tinkey}} \emph {et~al.},\ }\href {\doibase
  http://dx.doi.org/10.1063/1.4883638} {\bibfield  {journal} {\bibinfo
  {journal} {Appl. Phys. Lett.}\ }\textbf {\bibinfo {volume} {104}},\ \bibinfo
  {eid} {232410} (\bibinfo {year} {2014})}\BibitemShut {NoStop}%
\bibitem [{\citenamefont {Song}\ and\ \citenamefont
  {Dery}(2014)}]{Song_arxiv2014}%
  \BibitemOpen
  \bibfield  {author} {\bibinfo {author} {\bibfnamefont {Y.}~\bibnamefont
  {Song}}\ and\ \bibinfo {author} {\bibfnamefont {H.}~\bibnamefont {Dery}},\
  }\href {http://arxiv.org/abs/1401.7649} {} (\bibinfo {year} {2014}),\ \Eprint
  {http://arxiv.org/abs/1401.7649} {arXiv:1401.7649} \BibitemShut {NoStop}%
\bibitem [{\citenamefont {Ando}\ \emph {et~al.}(2011)\citenamefont {Ando} \emph
  {et~al.}}]{Ando_APL2011}%
  \BibitemOpen
  \bibfield  {author} {\bibinfo {author} {\bibfnamefont {Y.}~\bibnamefont
  {Ando}} \emph {et~al.},\ }\href {\doibase
  http://dx.doi.org/10.1063/1.3643141} {\bibfield  {journal} {\bibinfo
  {journal} {Appl. Phys. Lett.}\ }\textbf {\bibinfo {volume} {99}},\ \bibinfo
  {eid} {132511} (\bibinfo {year} {2011})}\BibitemShut {NoStop}%
\bibitem [{\citenamefont {Ando}\ \emph {et~al.}(2012)\citenamefont {Ando} \emph
  {et~al.}}]{Ando_PRB2012}%
  \BibitemOpen
  \bibfield  {author} {\bibinfo {author} {\bibfnamefont {Y.}~\bibnamefont
  {Ando}} \emph {et~al.},\ }\href {\doibase 10.1103/PhysRevB.85.035320}
  {\bibfield  {journal} {\bibinfo  {journal} {Phys. Rev. B}\ }\textbf {\bibinfo
  {volume} {85}},\ \bibinfo {pages} {035320} (\bibinfo {year}
  {2012})}\BibitemShut {NoStop}%
\bibitem [{\citenamefont {Ishikawa}\ \emph {et~al.}(2012)\citenamefont
  {Ishikawa} \emph {et~al.}}]{Ishikawa_APL2012}%
  \BibitemOpen
  \bibfield  {author} {\bibinfo {author} {\bibfnamefont {M.}~\bibnamefont
  {Ishikawa}} \emph {et~al.},\ }\href {\doibase
  http://dx.doi.org/10.1063/1.4728117} {\bibfield  {journal} {\bibinfo
  {journal} {Appl. Phys. Lett.}\ }\textbf {\bibinfo {volume} {100}},\ \bibinfo
  {eid} {252404} (\bibinfo {year} {2012})}\BibitemShut {NoStop}%
\bibitem [{\citenamefont {Hamaya}\ \emph {et~al.}(2013)\citenamefont {Hamaya}
  \emph {et~al.}}]{Hamaya_JAP2013}%
  \BibitemOpen
  \bibfield  {author} {\bibinfo {author} {\bibfnamefont {K.}~\bibnamefont
  {Hamaya}} \emph {et~al.},\ }\href {\doibase
  http://dx.doi.org/10.1063/1.4793501} {\bibfield  {journal} {\bibinfo
  {journal} {J. Appl. Phys.}\ }\textbf {\bibinfo {volume} {113}},\ \bibinfo
  {eid} {17C501} (\bibinfo {year} {2013})}\BibitemShut {NoStop}%
\bibitem [{\citenamefont {Kameno}\ \emph {et~al.}(2014)\citenamefont {Kameno}
  \emph {et~al.}}]{Kameno_APL2014}%
  \BibitemOpen
  \bibfield  {author} {\bibinfo {author} {\bibfnamefont {M.}~\bibnamefont
  {Kameno}} \emph {et~al.},\ }\href {\doibase
  http://dx.doi.org/10.1063/1.4867650} {\bibfield  {journal} {\bibinfo
  {journal} {Appl. Phys. Lett.}\ }\textbf {\bibinfo {volume} {104}},\ \bibinfo
  {eid} {092409} (\bibinfo {year} {2014})}\BibitemShut {NoStop}%
\bibitem [{\citenamefont {Yu}\ and\ \citenamefont
  {Flatt\'e}(2002)}]{Yu_PRB2002}%
  \BibitemOpen
  \bibfield  {author} {\bibinfo {author} {\bibfnamefont {Z.}~\bibnamefont
  {Yu}}\ and\ \bibinfo {author} {\bibfnamefont {M.}~\bibnamefont {Flatt\'e}},\
  }\href {\doibase 10.1103/PhysRevB.66.201202} {\bibfield  {journal} {\bibinfo
  {journal} {Phys. Rev. B}\ }\textbf {\bibinfo {volume} {66}},\ \bibinfo
  {pages} {201202} (\bibinfo {year} {2002})}\BibitemShut {NoStop}%
\bibitem [{\citenamefont {Jansen}\ and\ \citenamefont
  {Min}(2007)}]{Jansen_PRL2007}%
  \BibitemOpen
  \bibfield  {author} {\bibinfo {author} {\bibfnamefont {R.}~\bibnamefont
  {Jansen}}\ and\ \bibinfo {author} {\bibfnamefont {B.}~\bibnamefont {Min}},\
  }\href {\doibase 10.1103/PhysRevLett.99.246604} {\bibfield  {journal}
  {\bibinfo  {journal} {Phys. Rev. Lett.}\ }\textbf {\bibinfo {volume} {99}},\
  \bibinfo {pages} {246604} (\bibinfo {year} {2007})}\BibitemShut {NoStop}%
\bibitem [{\citenamefont {Harrison}(1961)}]{Harrison_PR1961}%
  \BibitemOpen
  \bibfield  {author} {\bibinfo {author} {\bibfnamefont {W.~A.}\ \bibnamefont
  {Harrison}},\ }\href {\doibase 10.1103/PhysRev.123.85} {\bibfield  {journal}
  {\bibinfo  {journal} {Phys. Rev.}\ }\textbf {\bibinfo {volume} {123}},\
  \bibinfo {pages} {85} (\bibinfo {year} {1961})}\BibitemShut {NoStop}%
\end{thebibliography}
%

\end{document}